\documentstyle[aps,prb,epsfig,twocolumn,floats]{revtex}

\def\(#1){(\ref{#1})}		

\begin{document} 
\draft
\def\dbltopfraction{1.0}

\wideabs{
\title{\bf Thermoelastic Damping in Micro- and\\ Nano-Mechanical
Systems} 
\author{Ron Lifshitz and M.~L.~Roukes}
\address{Condensed Matter Physics 114-36, California Institute of
Technology, Pasadena, CA 91125}
\date{\today}
\maketitle

\begin{abstract}
The importance of thermoelastic damping as a fundamental dissipation
mechanism for small-scale mechanical resonators is evaluated in light
of recent efforts to design high-Q micrometer- and nanometer-scale
electro-mechanical systems (MEMS and NEMS). The equations of linear
thermoelasticity are used to give a simple derivation for
thermoelastic damping of small flexural vibrations in thin beams. It
is shown that Zener's well-known approximation by a Lorentzian with a
single thermal relaxation time slightly deviates from the exact
expression.
\end{abstract}
}

\section{Introduction}

Micro-electro-mechanical systems (MEMS) and more recently
nano-electro-mechanical systems (NEMS) are being developed aggressively
for a variety of applications as well as for accessing
new regimes of basic experimental research. Among the different
applications envisioned for MEMS and NEMS are ultrafast and high precision
actuators, sensors (such as accelerometers, bolometers, magnetometers,
and calorimeters), and narrowband high-frequency mechanical filters,
all with compact and low-power designs that can be fully integrated with
modern semiconductor electronics. Experimentally, it is hoped that
NEMS will open the door to the investigation of new regimes of
phonon-mediated processes as well as the quantum behavior of
mesoscopic mechanical systems.\cite{rouk,cle}

For all these pursuits it is desired to design and construct systems
with very little loss of energy or very high quality factors
$Q$. Unfortunately, it has been consistently observed that the quality
factors of resonators decrease with size significantly---even when
made from pure single-crystal materials. It is therefore of great
importance to understand the dominating energy dissipation mechanisms
in mechanical resonators when one approaches submicron scales. Within
these mechanisms one would like to identify those which are
fundamental and always impose an upper limit on the quality factor and
those which might be eliminated through improved design and
fabrication.

In this work we examine the importance of the process of thermoelastic
damping as a dissipation mechanism in MEMS and NEMS. Although there
has been some discussion of different dissipation mechanisms in
MEMS\cite{ros,blom,mihail,grey,carr,yaspaper,yas} very few authors
have addressed the question of thermoelastic damping.
Roszhart\cite{ros} observed thermoelastic damping in single-crystal
silicon micro-resonators at room temperature; and Yasumura {\it et
al.}\cite{yas} recently reported thermoelastic damping in
micro-resonators an order of magnitude smaller than Roszhart's in
silicon nitride, also at room temperature. The question arises whether
one should be surprised by such observations or whether one should
expect to see thermoelastic damping at these length scales. We
establish in this paper that as long as the system remains in the
regime of ``diffusive thermal phonons'' (to be discussed below) the
latter is the case for two basic reasons:
\begin{enumerate}
\item The strength of the damping caused by thermoelastic coupling is
independent of geometry. It depends only on the thermodynamic
properties of the material as a function of temperature.

\item In the case of flexural vibrations of thin beams the position of
peak damping as a function of frequency depends on the dimensions of
the beam. Therefore, even though the normal frequencies of the
resonators increase as they become smaller so does the frequency at
which peak damping occurrs. 
\end{enumerate}
\noindent
These effects conspire together to maintain the relevance of
thermoelastic damping all the way down to the nanometer scale.

In the next section we describe the process of thermoelastic damping,
review some of the relevant literature, and present the outline of
this paper.

\section{The process of thermoelastic damping}
\label{tedamp}

Acoustic modes---such as a sound wave traveling through an infinitely
large elastic material or a normal mode of vibration of an elastic
resonator of finite geometry---will experience damping due to their
nonlinear interaction with a surrounding bath of {\it
thermally-excited\/} elastic modes, or phonons. If the mean free path
of these thermal phonons is much smaller than the wavelength of the
acoustic mode then sufficient thermalization occurrs on the scale of
interest. It is then possible to define a temperature locally, even
when the system is not in a state of thermal
equilibrium. Equivalently, if the relaxation rate of the phonon
distribution to a local Bose-Einstein distribution is much faster than
the frequency of the acoustic mode then one has a well-defined
temperature field, and there is no need to treat the thermal phonons
as individual excitations.  In this regime, in which thermal phonons
are said to be ``diffusive,'' the complicated interaction between the
acoustic mode and the thermal phonon bath in an isotropic solid is
captured by a single macroscopic parameter---the material's thermal
expansion coefficient
\begin{equation}\label{alpha}
\alpha={1\over L}{\partial L\over\partial T}\ ,
\end{equation}
which couples changes of length to changes of temperature. Note that
here and throughout the paper we make use of the {\it linear\/}
coefficient of thermal expansion whose value is one third of the
volumetric coefficient of thermal expansion.

When an elastic solid is set in motion it is taken out of equilibrium,
having an excess of kinetic and potential energy. In an isothermal and
perfectly-linear {\it elastic\/} solid such a non-equilibrium state
can exist forever.  In a {\it thermoelastic\/} solid the coupling of
the strain field to a temperature field provides an energy dissipation
mechanism which allows the system to relax back to equilibrium.
Relaxation of the thermoelastic solid is achieved through the
irreversible flow of heat driven by local temperature
gradients that through the coupling accompany the strain field. This
process of energy dissipation, called {\it thermoelastic damping,} is
a fundamental one. As long as $\alpha$---which acts as a coupling
constant---is non-zero, thermoelastic damping introduces an upper
limit to the quality factor of even the most perfectly designed and
constructed resonator.

The first to realize that thermoelastic relaxation may be a
significant source of damping in mechanical resonators was Zener, who
in a series of papers\cite{zenpapers} in the 1930's developed a
general theory of thermoelastic damping. Zener treated the problem in
the framework of his so-called ``standard model'' of the anelastic
solid\cite{zenbook} and showed that the damping behavior can be
approximated very well by a single relaxation peak with a
characteristic relaxation time.  This relaxation time corresponds to
the thermal diffusion time across the width $b$ of the beam, which is
proportional to $b^2/\chi$, where $\chi$ is the solid's thermal
diffusivity [to be defined in Sec.~\ref{te} Eq.~\(heat)].  We shall
review Zener's theory in section~\ref{zenersection}.

It was only two decades later that other researchers began reexamining
the question of thermoelastic damping by seeking exact solutions to
the coupled equations of linear thermoelasticity in various
geometries. The solutions for propagating plane waves in an infinite
thermoelastic solid\cite{biot,der,chad,chadreview} showed that the two
transverse modes, which propagate through the solid without causing
any local volume changes, do not couple to the temperature field and
hence suffer no damping. The longitudinal mode, on the other hand,
does couple to the temperature field and an exact expression has been
obtained for its attenuation and dispersion. Solutions for
thermoelastic Rayleigh waves (two-dimensional surface waves on a
semi-infinite solid) immediately followed,\cite{lock} but progress on
thermoelastic solids with finite geometries came much later due to the
well-known difficulty of solving even the equations of linear
elasticity with finite boundary conditions.\cite{green} Nevertheless,
analytic solutions now exist, at least in terms of the dispersion
relations, for thermoelastic waves in an infinite thin
plate,\cite{daimplate} and longitudinal waves in infinite rods with
circular cross sections.\cite{chadreview,daimrod,suhubi} Other
geometries, such as beams of rectangular cross sections, have been too
difficult to solve analytically. To treat such problems one generally
needs to use approximate theories.

MEMS and NEMS resonators generally contain elements which vibrate in
either torsional or flexural modes. Because of the way one fabricates
such devices, rectangular cross sections often turn out to be the most
relevant.  Pure torsional modes of rectangular beams involve no local
volume changes and therefore, just as for transverse waves in the
bulk, they do not suffer any thermoelastic losses. For flexural
vibrations of thin rectangular beams one may use
Zener's\cite{zenpapers} approximate expression
[Sec.~\ref{zenersection} Eq.~\(zenerq)] for thermoelastic
damping. This appears to be the general practice.\cite{ros,yas} On the
other hand, one may try to seek exact solutions of the thermoelastic
equations for the case of a {\it thin\/} beam under flexure.  Landau
and Lifshitz\cite{lan} have provided an exact expression for the
attenuation coefficient of such vibrations without, however, giving a
rigorous derivation and solution of the equations. Shieh\cite{shieh}
investigated the thermoelastic beam equations in a more general
context while performing a dynamic instability analysis of
vertically-standing cantilevers. He solved the equations, giving a
plot of thermoelastic damping $Q^{-1}$ in thin rectangular beams, but
fell short of providing an analytic expression for $Q^{-1}$.

To remedy this state of affairs we give, in section~\ref{te}, a simple
derivation of the approximate thermoelastic equations for a thin beam
under flexure, and then solve these equations in
section~\ref{solution} to arrive at an exact expression
[Sec.~\ref{solution} Eq.~\(qinv)] for thermoelastic damping in thin
rectangular beams. Although Zener's approximation is good for many
situations we suggest that in the future the exact expression given
here be used instead. In section~\ref{exp} we discuss the experimental
implication of our results showing the characteristic damping curves
expected for GaAs and Silicon, which are typical materials used in the
fabrication of micrometer- and nanometer-scale resonators. In
section~\ref{longi} we say a few words about thermoelastic damping of
longitudinal waves in MEMS and NEMS, and in section~\ref{conc} we
conclude by discussing the validity of the theory which we present
here. In the appendix we give a careful analysis comparing the exact
expression for thermoelastic damping in thin beams with the
approximate Lorentzian behavior, predicted by Zener.

\section{Zener's standard model of the linear anelastic solid}
\label{zenersection}

Zener's standard model of anelasticity~\cite{zenbook} is based on an
extension of Hooke's law to the most general linear homogeneous
equation involving stress $\sigma$, strain $\epsilon$, and their first
time derivatives:
\begin{equation}\label{linearsolid}
\sigma +\tau_\epsilon \dot\sigma= M_R(\epsilon + \tau_\sigma\dot\epsilon).
\end{equation}
The physical interpretation of the three parameters of the model is
straightforward.  When the strain is kept constant the stress relaxes
exponentially with a relaxation time $\tau_\epsilon$. Similarly,
$\tau_\sigma$ is the strain relaxation time when the stress is kept
constant. $M_R$ is the value of the pertinent elastic modulus after
all relaxation has occurred. $M_U = M_R (\tau_\sigma/\tau_\epsilon)$ is
the unrelaxed value of the elastic modulus.

Under periodic dynamical conditions  
\begin{equation}\label{periodic}
\sigma(t)=\sigma_0 e^{i\omega t},\qquad 
\epsilon(t)=\epsilon_0 e^{i\omega t},
\end{equation}
the stress and strain amplitudes are related by a frequency-dependent
complex elastic modulus.  The dissipation, or ``internal friction,''
in the solid $Q^{-1}$ is defined as the fraction of energy lost per
radian of vibration.  If it is small, the dissipation is equal to the
ratio of the imaginary and real parts of the complex modulus, giving
\begin{equation}\label{zenermodel}
Q^{-1} = \Delta_M{\omega\tau\over{1+(\omega\tau)^2}}\ ,
\end{equation}
where $\tau=\sqrt{\tau_\sigma\tau_\epsilon}$, and 
\begin{equation}\label{relaxation}
\Delta_M = {{M_U-M_R}\over \sqrt{M_R M_U}}
\end{equation} is a dimensionless quantity called the ``relaxation
strength'' of the modulus. 

Thus, the dissipation exhibits a Lorentzian behavior as a function of
$\omega\tau$ with a maximum value of $\Delta_M/2$ when $\omega\tau=1$.
Dissipation peaks of this form, generally called ``Debye peaks,'' are
quite ubiquitous as to be expected from their prediction by such a
na\"{\i}ve model. They occur for many different relaxation mechanisms
such as point-defect relaxation (``Snoek peaks''), defect pair
reorientation (``Zener peaks''), dislocation relaxation (``Bordoni
peaks''), grain boundary relaxation, and of course thermal
relaxation.\cite{now} In many of these examples there is not just a
single relaxation time $\tau$ and therefore one sees multiple or
broadened Debye peaks.

One can understand qualitatively why there is a peak in dissipation
when $\omega\tau=1$ in the following way: If the frequency of
vibration $\omega$ is much smaller than the effective relaxation rate
$1/\tau$ of the solid then the system remains essentially in
equilibrium and very little energy is dissipated. If the vibration
frequency is much larger than the effective relaxation rate
$\omega\gg1/\tau$, the system has no time to relax and again very
little energy is dissipated. It is only when the vibration frequency
is on the order of the system's effective relaxation rate that
appreciable dissipation occurs. The full picture may be more
complicated, however, because in some cases $\tau$ itself can depend
on $\omega$.

In the case of a thermoelastic solid the relaxation strength
\(relaxation) to be considered is that of Young's modulus
\begin{equation}\label{epsilon}
\Delta_E = {{E_{ad}-E}\over E} = {E\alpha^2 T_0\over C_p} ,
\end{equation}
whose value is known from basic thermodynamics. Here $E_{ad}$ is the
unrelaxed, or {\it adiabatic,} value of Young's modulus and $E$ is its
relaxed, or {\it isothermal,} value.  $C_p$ is the heat capacity per
unit volume at constant pressure, or stress, but replacing it by the
heat capacity $C_v$ at constant volume, or strain, will introduce an
error in $\Delta_E$ which is only on the order of ${\Delta_E}^2$.
Since $Q^{-1}$, and therefore also $\Delta_E$, are assumed small such
an error is negligible. For similar considerations, no harm is done by
replacing $\sqrt{E_{ad} E}$ in the denominator of \(relaxation) by $E$.

Zener\cite{zenpapers,zenbook} calculated the thermal relaxation times
associated with different transverse thermal modes for a thin beam
under flexure. He showed that for rectangular beams approximately
98.6$\%$ of the relaxation occurs through the first mode whose
relaxation time is
\begin{equation}\label{zenertau}
\tau_{\rm z} = {b^2\over{\pi^2\chi}}\ ,
\end{equation}
where $\chi$ is the thermal
diffusivity of the solid and $b$ is the width of the beam. Only a very
small error is therefore made by considering the vibrating thin beam
as having a simple relaxation \(zenermodel) with a single relaxation time
$\tau_{\rm z}$,
\begin{equation}\label{zenerq}
Q^{-1}_{\rm Z} = {E\alpha^2 T_0\over C_p}
{\omega\tau_{\rm z}\over{1+(\omega\tau_{\rm z})^2}}\ .
\end{equation}
Nevertheless, we show below that for the simple geometry of a
rectangular beam---which is the most relevant for many current MEMS and
NEMS designs---such an expansion in transverse thermal eigenmodes is
unnecessary and an {\it exact\/} expression for thermoelastic damping
can easily be obtained.

\section{Thermoelastic equations of a thin beam}
\label{te}

We consider small flexural displacements of a thin elastic beam of
length $L$ and rectangular cross section of dimensions $b\times c$. We
define the $x$-axis along the axis of the beam and the $y$- and
$z$-axes parallel to the surfaces of dimensions $b$ and $c$,
respectively. In equilibrium, the beam is unstrained, unstressed, and
at temperature $T_0$ everywhere. Departure of the beam from
equilibrium is described by a displacement field $u_i$ ($i=x,y,z$) and
a temperature field $T=T_0+\theta$. The displacement field $u_i$ and
the relative temperature field $\theta$, as well as the strain and
stress tensors $u_{ij}$ and $\sigma_{ij}$, are all functions of
position and time.

We consider pure transverse motion $Y(x)$ in the $y$ direction and make
the usual Euler-Bernoulli assumption that the transverse dimensions of
the beam, $b$ and $c$, are sufficiently small compared with the length
$L$ of the beam and the radius of curvature $R$ of the bending that
any plane cross section, initially perpendicular to the axis of the
beam, remains plane and perpendicular to the neutral surface during
bending. The neutral surface is the one running through the length of
the beam which suffers no extension or contraction during its bending.

We take the surfaces of the beam to be stress free which implies that
all but the $\sigma_{xx}$ component of the stress tensor vanish on the
surface. Because the beam is thin this approximately holds in its
interior as well. Hooke's law for the thermoelastic beam then takes a 
rather simple form
\begin{mathletters}
\begin{eqnarray}\label{hooke}
u_{xx}&=&{1\over E}\sigma_{xx} + \alpha\theta, \\
u_{yy}&=&u_{zz}=-{\sigma\over E}\sigma_{xx} + \alpha\theta, \\ 
u_{xy}&=&u_{yz}=u_{zx}=0, 
\end{eqnarray}
\end{mathletters}
where $E$ is Young's modulus, $\sigma$ is Poisson's ratio, $\alpha$ is
the linear thermal expansion coefficient \(alpha), and we have taken
into account the fact that strain arises both from mechanical stress
as well as thermal expansion. It is simple to show (see, for example,
Landau and Lifshitz\cite{lan}) that the longitudinal strain component
$u_{xx}$, a distance $y$ away from the neutral surface, is equal to
$y/R$. By replacing the curvature of the beam $1/R$ with $-\partial^2
Y/\partial x^2$ we may express the non-zero components of the strain
field in the beam as
\begin{mathletters}\label{strain}
\begin{eqnarray}
u_{xx}&=&-y{\partial^2 Y\over\partial x^2}, \\
u_{yy}&=&u_{zz}=\sigma y{\partial^2 Y\over\partial x^2} 
+ (1+\sigma)\alpha\theta.
\end{eqnarray}
\end{mathletters}

Following the standard derivation procedure for isothermal beams with
no thermoelastic coupling\cite{lan} but with the modified
thermoelastic strain \(strain) leads to an equation of motion for the
beam of the form
\begin{equation}\label{beameqn}
\rho A{\partial^2 Y\over\partial t^2} +
{\partial^2\over\partial x^2}\biggl(EI{\partial^2 Y\over\partial x^2}
+ E\alpha I_T\biggr) = 0\ ,
\end{equation}
where $\rho$ is the density of the beam, and $A=bc$ is the area of its
cross section. The quantities $I$ and $I_T$ are integrals over the
cross section of the beam giving the mechanical and the thermal
contributions to its moment of inertia
\begin{equation}\label{inertia}
I=\int_A y^2 dy dz = {b^3 c\over 12},\quad {\rm and}\quad  
I_T=\int_A y\theta dy dz.
\end{equation}
In evaluating the moment of inertia $I$ we have neglected the
deviation of the cross section from its rectangular shape which arises
from having a non-zero Poisson ratio $\sigma$. Such an approximation
is justified for small deflections since the error it introduces is
only on the order of the transverse beam dimension divided by the
radius of curvature of the bending.

To the equation of motion \(beameqn) we add the heat equation, which
in the presence of thermoelastic coupling is given by\cite{lan}
\begin{equation}\label{heat}
{\partial \theta\over\partial t} = \chi \nabla^2 \theta - 
{E\alpha T\over{(1-2\sigma)C_v}} {\partial\over\partial t}
\sum_j u_{jj}\ .
\end{equation}
We make two simplifications to this equation. First, since $\theta\ll
T_0$, we can safely replace $T$ by $T_0$ in the second term on the
right-hand side of the equation. Not doing so will introduce
unnecessary nonlinearities into the problem. Second, noting that
thermal gradients in the plane of the cross section along the
$y$-direction are much larger than gradients along the beam axis, and
that no gradients exist in the $z$-direction, we replace
$\nabla^2\theta$ by $\partial^2 \theta/\partial y^2$. Substituting the
value of the strain field from Eq. \(strain) we finally get
\begin{equation}\label{finalheat}
\biggl(1+2\Delta_E{1+\sigma\over{1-2\sigma}}\biggr){\partial
\theta\over\partial t} =
\chi {\partial^2 \theta\over\partial y^2} +
y{\Delta_E\over\alpha}{\partial\over\partial t}{\partial^2
Y\over\partial x^2}\ ,
\end{equation}
where we have identified the relaxation strength of Young's modulus
$\Delta_E$ \(epsilon).

\section{Solution of the thermoelastic equations for harmonic vibrations} 
\label{solution}

\begin{figure}[t]
\begin{center}
\epsfig{file=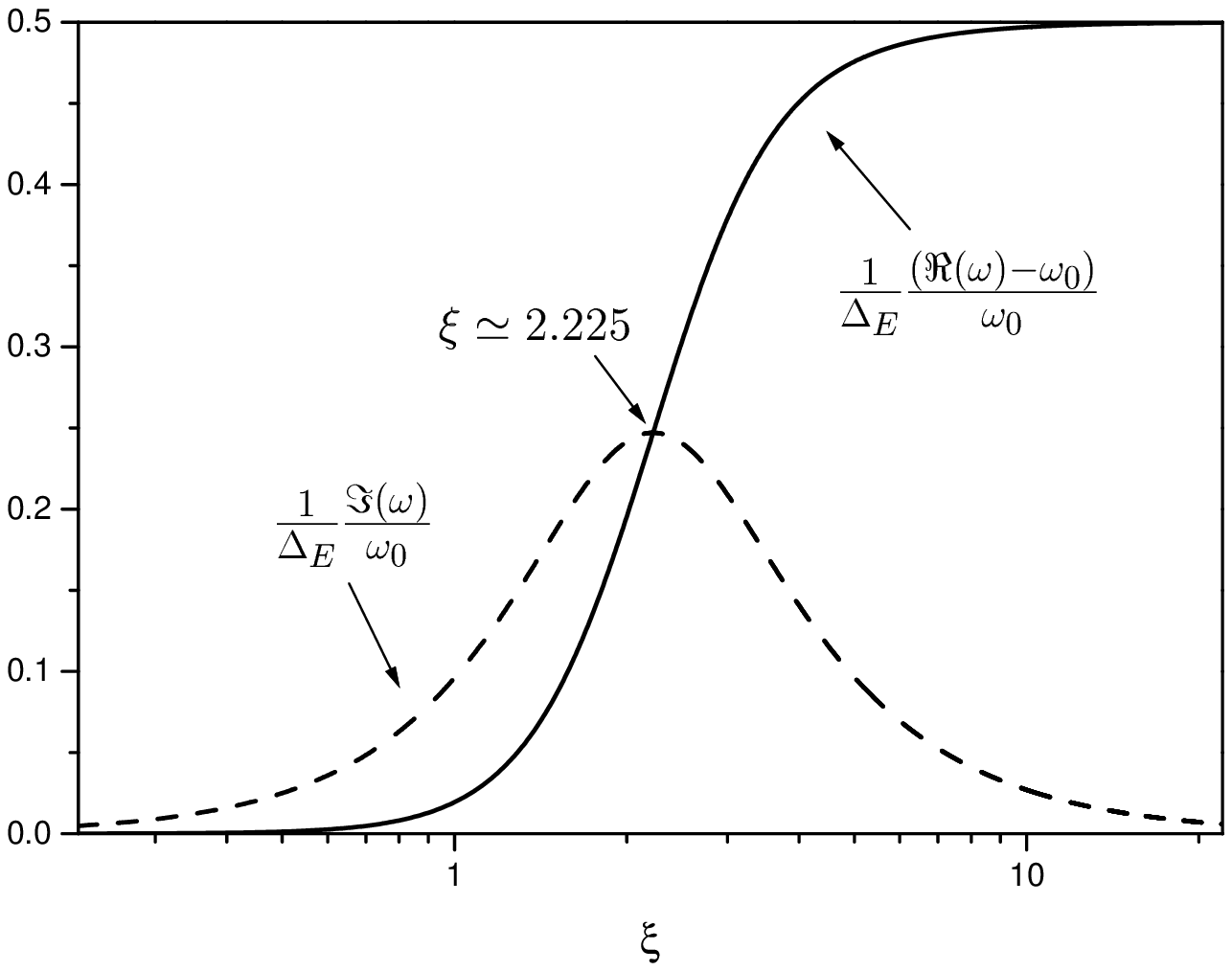,width=\columnwidth}
\end{center}
\caption{Universal plots of the frequency shift and attenuation,
Eq.\(ReandIm), of small flexural vibrations in thin rectangular beams
due to thermoelastic coupling. }
\label{ReandImplot}
\end{figure}

\begin{figure*}[bt]
\begin{center}
\epsfig{file=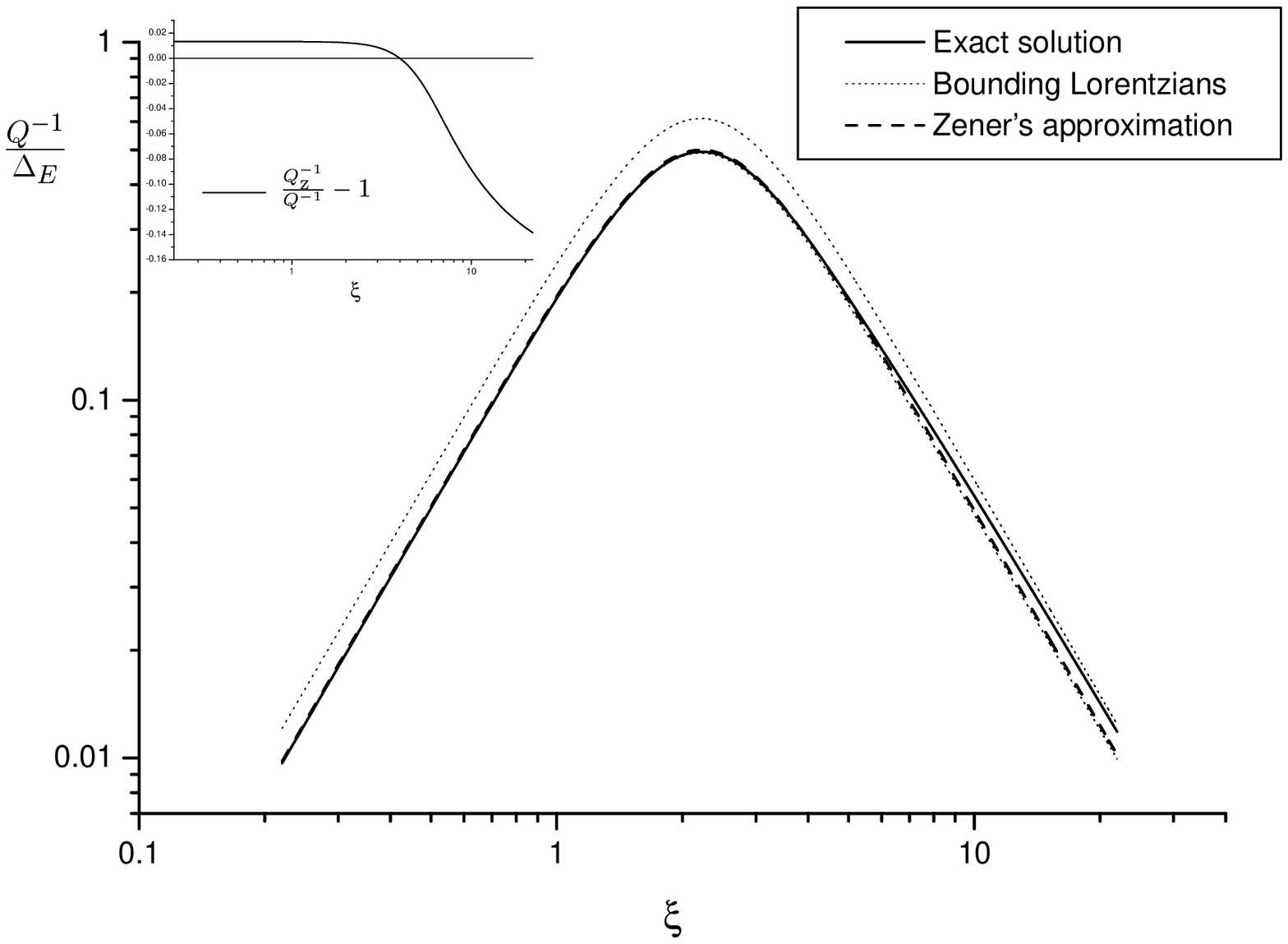,width=5.0in}
\end{center}
\caption{Universal plot of thermoelastic damping of small flexural
vibrations in thin beams \(qinv). The damping is ploted in units of
the relaxation strength $\Delta_E=E\alpha^2 T_0/C$, as a function of
the dimensionless variable $\xi=b\sqrt{\omega_0/2\chi}$ for one decade
above and below its peak value $\simeq 0.494$, occurring at $\xi\simeq
2.225$.  The two bounding Lorentzians \(bounds) in the variable
$\xi^2/\sqrt{24}$ are shown along with Zener's approximation
$Q^{-1}_{\rm Z}/\Delta_E$ \(zenerq). The relative error in Zener's
approximation $(Q^{-1}_{\rm Z}-Q^{-1})/Q^{-1}$ is shown in the inset.}
\label{universalplot}
\end{figure*}

To calculate the effect of thermoelastic coupling on the vibrations of
a thin beam we solve the coupled thermoelastic equations \(beameqn) and
\(finalheat) for the case of harmonic vibrations. We set
\begin{equation}\label{Yoft}
Y(x,t)=Y_0(x)e^{i\omega t},
\quad\theta(x,y,t)=\theta_0(x,y)e^{i\omega t},
\end{equation}
calculate the temperature profile along the beam's cross section using
the heat equation \(finalheat), and substitute it into the equation of
motion \(beameqn) in order to obtain the normal modes of vibration and
their corresponding frequencies. We expect to find that in general the
frequencies are complex, the real part $\Re(\omega)$ giving the new
eigenfrequencies of the beam in the presence of thermoelastic
coupling, and the imaginary part $|\Im(\omega)|$ giving the
attenuation of the vibration. The amount of thermoelastic damping,
expressed in terms of the inverse of the quality factor, will then be
given by
\begin{equation}\label{qinvdef}
Q^{-1} = 2\biggl|{\Im(\omega)\over \Re(\omega)}\biggr|,
\end{equation}
which is the fraction of energy lost per radian, the factor of 2
arising from the fact that the mechanical energy of the beam is
proportional to the square of its amplitude.

Substituting \(Yoft) into the heat equation \(finalheat) and
neglecting the term of order $\Delta_E$ on its left-hand side (which
will only introduce a correction of order $\Delta_E^2$ to the final
result) yields the following equation for $\theta_0$
\begin{equation}\label{heat1}
{\partial^2 \theta_0\over\partial y^2} =
i{\omega\over\chi}\biggl(\theta_0 - {\Delta_E\over\alpha}{\partial^2
Y_0\over\partial x^2} y\biggr)\ ,
\end{equation}
whose solution is
\begin{equation} \label{thetasolution}
\theta_0 - {\Delta_E\over\alpha}{\partial^2 Y_0\over\partial x^2} y
= A\sin(k y) + B\cos(k y)\ ,
\end{equation}
where
\begin{equation}\label{k}
k = \sqrt{i{\omega\over\chi}} = 
(1+i) \sqrt{\omega\over{2\chi}}\ .
\end{equation}
The coefficients $A$ and $B$ are determined by taking as boundary
conditions the requirement that there be no flow of heat across the
boundaries of the beam so that $\partial\theta_0/\partial y = 0$ at
$y=\pm b/2$. The temperature profile across the beam is then given by
\begin{equation}\label{profile}
\theta_0(x,y) = {\Delta_E\over\alpha}{\partial^2 Y_0(x)\over\partial x^2}
\biggl(y - {\sin(ky)\over{k\cos({bk\over 2})}}\biggr)\ .
\end{equation}

Now that we have the temperature profile we can substitute it into the
integral $I_T$ \(inertia) for the cross section's thermal moment.
Because $E$, $I$, and $\Delta_E=E\alpha^2 T_0/C$ are all constant
along the beam, the beam equation may subsequently be expressed as
\begin{equation}\label{finalbeam}
\omega^2 Y_0 = {EI\over{\rho A}}\biggl[1+
\Delta_E\biggl(1+f(\omega)\biggr)\biggr] 
{\partial^4 Y_0\over{\partial x^4}} ,
\end{equation}
where the complex function $f(\omega)$ is given by
\begin{equation}\label{fofomega}
f(\omega) = f(k(\omega)) = {24\over{b^3k^3}}\biggl({bk\over2} -
\tan\biggl({bk\over2}\biggr)\biggr) .
\end{equation}

The equation of motion for the beam \(finalbeam) is formally identical
to that of the isothermal beam with no thermoelastic coupling. The
only difference being that the isothermal value of Young's modulus $E$
is replaced by a frequency dependent modulus
\begin{equation}\label{Young}
E_\omega = E\biggl[1+ \Delta_E\biggl(1+f(\omega)\biggr)\biggr] .
\end{equation}
When $\omega$ becomes very large, $f(\omega)\to 0$, and Young's
modulus tends to its adiabatic, or unrelaxed, value
$E_{ad}=E(1+\Delta_E)$. When $\omega$ is very small, $f(\omega)\to
-1$, and Young's modulus recovers its isothermal value $E$, as
expected. For intermediate frequencies $E_\omega$ is complex.

The normal modes of vibration of the beam are given, as in the
isothermal case, by
\begin{equation}\label{Ysolution}
Y_0(x)=A\sin(qx) + B\cos(qx) + C\sinh(qx) + D\cosh(qx) ,
\end{equation}
where the coefficients $A$ through $D$ and the allowed values of $q$
are determined, as usual, by the boundary conditions at the two ends
of the beam. For example, for beams clamped at both ends or free at
both ends $q_nL=a_n=\{4.730, 7.853, 10.996, \ldots\}$, and for cantilevers
clamped at one end and free at the other $q_nL=a_n=\{1.875, 4.694, 7.855,
\ldots\}$, in all three cases tending for large $n$ to odd-integer
multiples of $\pi/2$. The dispersion relation between $\omega$ and $q_n$
for the thermoelastic beam is given by
\begin{equation}\label{dispersion}
\omega = \sqrt{E_\omega I\over{\rho A}} q_n^2
=\omega_0 \sqrt{1+ \Delta_E\biggl(1+f(\omega)\biggr)} ,
\end{equation}
where $\omega_0$ is the isothermal value of the eigenfrequency.

Neglecting corrections of order $\Delta_E^2$ we may replace $f(\omega)$
in the square root by $f(\omega_0)$. The dispersion relation
\(dispersion) then becomes
\begin{equation}\label{finaldispersion}
\omega = \omega_0
\biggl[1+ {\Delta_E\over 2}\biggl(1+f(\omega_0)\biggr)\biggr],
\end{equation}
from which we can easily extract the real and the imaginary parts,
giving the thermoelastic corrections of order $\Delta_E$ to the
eigenfrequencies of the beam as well as the corresponding attenuation
coefficients
\begin{mathletters}\label{ReandIm}
\begin{eqnarray}
&\Re(\omega) &= \omega_0 \biggl[1+ {\Delta_E\over 2}\biggl(1-
{6\over{\xi^3}} {\sinh\xi - \sin\xi\over{\cosh\xi + \cos\xi}}
\biggr)\biggr],\label{Reomega} \\[6pt]
&\Im(\omega) &= \omega_0
{\Delta_E\over 2} \biggl({6\over{\xi^3}}{\sinh\xi +
\sin\xi\over{\cosh\xi + \cos\xi}}-{6\over{\xi^2}}\biggr),
\label{Imomega}
\end{eqnarray}
\end{mathletters}
where 
\begin{equation}\label{xi}
\xi=b\sqrt{\omega_0\over{2\chi}}.
\end{equation}
The universal behavior of the normalized frequency shift
$(\Re(\omega)-\omega_0)/\omega_0\Delta_E$ and of the normalized
attenuation $\Im(\omega)/\omega_0\Delta_E$ as functions of the
dimensionless variable $\xi$ are shown in Figure~\ref{ReandImplot}.

Using the definition \(qinvdef) of the quality factor we arrive at an
expression for thermoelastic damping in a thin beam which, to first
order in $\Delta_E$, is given by
\begin{equation}\label{qinv}
Q^{-1} = {E\alpha^2 T_0\over C}\biggl({6\over{\xi^2}} -
{6\over{\xi^3}}{\sinh\xi + \sin\xi\over{\cosh\xi + \cos\xi}}\biggr).
\end{equation}

\begin{figure}[t]
\begin{center}
\epsfig{file=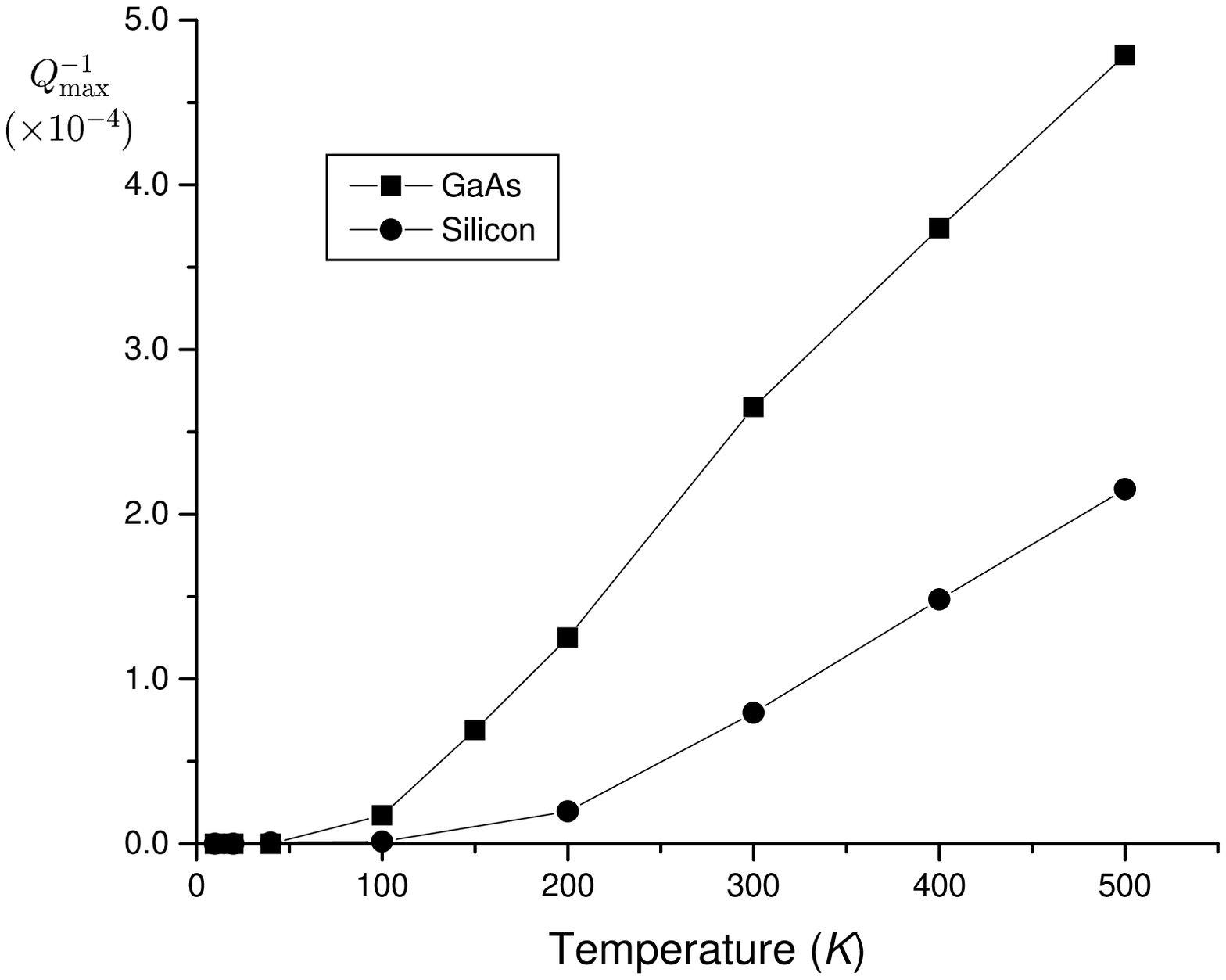,width=\columnwidth}
\end{center}
\caption{Peak value of thermoelastic damping $Q^{-1}_{\rm
max}$ $\simeq 0.494\Delta_E$ (maximum value of the plot in
Fig.~\ref{universalplot}) plotted for Gallium Arsenide and for Silicon
as a function of temperature.  Experimentally reported
values~\cite{gaas,sil} are used for the thermodynamic quantities $E$,
$\alpha$, and $C$. }
\label{strengthplot}
\end{figure}

\section{Discussion of the results and their experimental implications}
\label{exp}

We show in the Appendix that the exact expression \(qinv) for
thermoelastic damping is tightly bounded between two Lorentzians in
the variable $\xi^2/\sqrt{24}$, and that it behaves as $\xi^2$ for
small $\xi$ and as $1/\xi^2$ for large values of $\xi$. We also
compare the exact result with Zener's approximate expression
\(zenerq). The damping is peaked at $\xi_0\simeq 2.225$ with a maximum
value of $Q^{-1}_{\rm max}/\Delta_E \simeq 0.494$. The universal
behavior of $Q^{-1}/\Delta_E$ as a function $\xi$ is shown in
Figure~\ref{universalplot}.

The first conclusion to be drawn from this universal behavior is that
the peak value of thermoelastic damping, given approximately by
$0.494\Delta_E$, is independent of the dimensions of the beam. It only
depends on temperature through the thermodynamic properties $E$,
$\alpha$, and $C$ of the material. The values of $Q^{-1}_{\rm max}$
for GaAs and Silicon, typical materials used in the fabrication of
MEMS and NEMS, are plotted as a function of temperature in
Figure~\ref{strengthplot}. Thus, for example, at room temperature one
expects to observe quality factors no higher than $10^3$--$10^4$ if
one is operating at around $\xi=2.225$.

The dimensionless variable $\xi^2$ is proportional to the product
$\omega\tau$ used in Zener's model of the anelastic solid. Because we
are mainly concerned with the dependence of thermoelastic damping on
scale, and since both $\omega$, the beam's eigenfrequency, and
$\tau$, its thermal relaxation time, depend on the dimensions of the
beam, we find it more suitable to explicitly express the dependence of
$\xi$ on the dimensions of the beam. Instead of the usual plots of
Debye peaks as a function of frequency, we plot the damping curve as a
function of the dimensions of the beam.

To do that we express the (isothermal) eigenfrequencies \(dispersion)
in terms of the beam dimensions
\begin{equation}\label{eigen}
\omega^{(n)}=a_n^2 {b\over L^2} \sqrt{E\over{12\rho}}\ .
\end{equation}
We then get
\begin{equation}\label{xibeam}
\xi^2={a_n^2\over{4\sqrt{3}}}\ {b^3\over{L^2\ell_T}}\ ,
\end{equation}
where $\ell_T$ is a thermal diffusion length, proportional to the
phonon mean free path, given by
\begin{equation}\label{ellT}
\ell_T=\chi\sqrt{\rho\over E}.
\end{equation}

\begin{table}[t]
\begin{tabular}{lccc}
\multicolumn{4}{c}{\it GaAs}\\
\hline
$T$ & 10$K$ & 100$K$ & 300$K$\\
\hline
$\Delta_E(T)$ & 2.612 $10^{-8}$ & 1.718 $10^{-5}$ & 2.651 $10^{-4}$\\
$\ell_T$ ($\mu m$) & 1.300 $10^{+2}$ & 4.456 $10^{-2}$ & 6.455 $10^{-3}$\\
\hline\hline
\multicolumn{4}{c}{\it Silicon}\\
\hline
$T$ & 10$K$ & 100$K$ & 300$K$\\
\hline
$\Delta_E(T)$ & 2.319 $10^{-10}$ & 1.232 $10^{-6}$ & 7.942 $10^{-5}$\\
$\ell_T$ ($\mu m$) & 4.977 $10^{+2}$ & 2.017 $10^{-1}$ & 1.257 $10^{-2}$\\
\end{tabular}
\caption{Relaxation strengths $\Delta_E(T)$ and thermal diffusion
lengths $\ell_T$ (in micrometers) for GaAs and Silicon at three
representative temperatures.  The values are calculated from
experimental data,\cite{gaas,sil} and is used for generating
the plots of Figure~\ref{geometryplot}.}
\label{ellTable}
\end{table}

\begin{figure*}[p]
\begin{center}
\leavevmode
\hbox{%
\epsfig{file=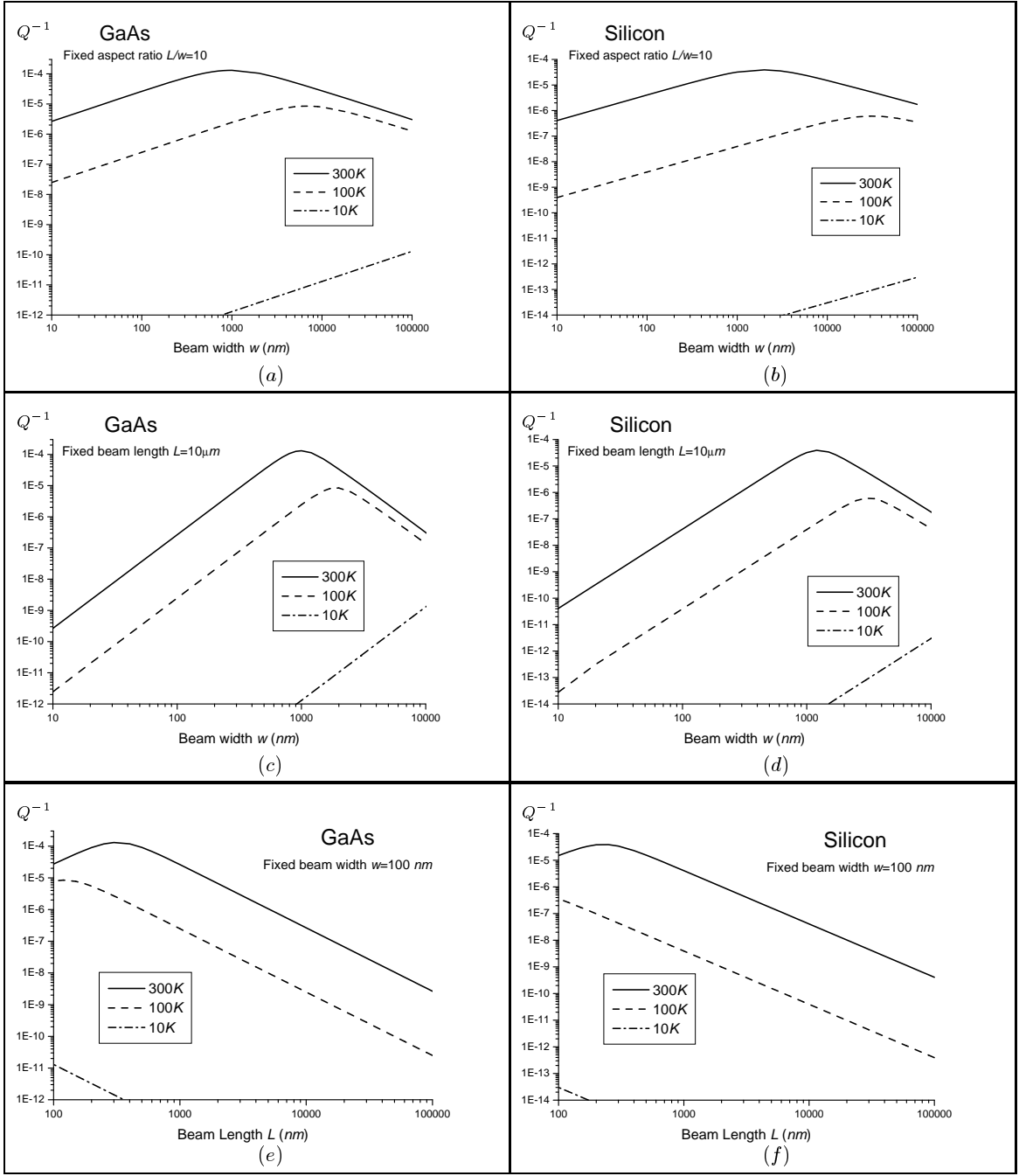,width=440pt}
}
\end{center}
\caption{Thermoelastic damping in Gallium Arsenide and Silicon thin
rectangular beams plotted for different geometries. In all cases the
beams are assumed clamped at both ends and vibrating at their
fundamental flexural mode.  The corresponding thermal diffusion
lengths $\ell_T$ and relaxation strengths $\Delta_E(T)$, used here,
are listed in Table~\ref{ellTable}. }
\label{geometryplot}
\end{figure*}

We use experimentally reported values of the thermodynamic properties
of GaAs\cite{gaas} and Silicon\cite{sil} to obtain $\ell_T$ as well
as $\Delta_E(T)$, listed in Table~\ref{ellTable}, for three
representative temperatures: 10$K$, 100$K$, and 300$K$.  We then use
these values for illustrative purposes to plot the dependence of
thermoelastic damping on geometry in three different ways:
\begin{enumerate}
\item $Q^{-1}$ {\it vs.} beam width $b$ for fixed aspect ratio $L/b$,
\item $Q^{-1}$ {\it vs.} beam width $b$ for fixed beam length $L$,
\item $Q^{-1}$ {\it vs.} beam length $L$ for fixed beam width $b$.
\end{enumerate}
The outcome is shown in Figure~\ref{geometryplot} for the case of a
beam clamped at both ends vibrating in its fundamental mode
($a_n=4.73$).  The slopes of the curves, plotted in log-log scale, are
$\pm1$, $\pm3$, and $\pm2$ respectively, which is easily understood
through the relation \(xibeam) and the fact that $Q^{-1}$ grows as
$\xi^2$ for small $\xi$ and decays as $\xi^{-2}$ for large $\xi$.

It is clear from these examples that thermoelastic damping is a
significant source of dissipation for MEMS and NEMS at temperatures
around 100$K$ and above. The reason for this is very simple and
follows from the fact that as the beam becomes smaller its
eigenfrequency increases at the same time that its thermal relaxation
time decreases. The product of the two, which can be controlled by
independently varying the two dimensions $b$ and $L$, can therefore
remain of order unity down to the nanometer scale.


\section{Thermoelastic damping of longitudinal waves}
\label{longi}

\begin{table}[t]
\begin{tabular}{lccc}
\multicolumn{4}{c}{\it GaAs}\\
\hline
$T$ & 10$K$ & 100$K$ & 300$K$\\
\hline
$\nu_l$ & 7.1 MHz & 10.0 GHz & 138 GHz\\
\hline\hline
\multicolumn{4}{c}{\it Silicon}\\
\hline
$T$ & 10$K$ & 100$K$ & 300$K$\\
\hline
$\nu_l$ & 3.1 MHz & 7.6 GHz & 121 GHz\\
\end{tabular}
\caption{Thermal relaxation rates $\nu_l=1/2\pi\tau_l=c_l^2/2\pi\chi$
for longitudinal thermoelastic waves in GaAs and Silicon at three
representative temperatures.  The values are calculated from
experimentally reported data,\cite{gaas,sil} and listed here to
illustrate the typical frequencies at which thermoelastic damping of
longitudinal waves is most significant. The relaxation strengths at
these temperatures are the same as those listed in
Table~\ref{ellTable}.  }
\label{longTable}
\end{table}

For completeness we would like to say a few words regarding
thermoelastic damping of longitudinal modes. Even though such modes do
not come into play when considering MEMS and NEMS resonators they may
affect general questions of heat flow and energy relaxation in other
elements of MEMS and NEMS. The process of thermoelastic damping of
longitudinal waves is similar to that of flexural waves in that energy
is dissipated through the irreversible flow of heat from hot to cold
regions of the solid. The difference is that the distance between
these regions is not fixed by the transverse geometry of the
device. Because a longitudinal wave is a compression wave hot and cold
regions are separated by half a wave length $\lambda$ along the
propagation direction of the wave. The wave length $\lambda$ is
inversely proportional to the frequency and therefore the thermal
relaxation time $\tau_l$ for a longitudinal thermoelastic wave, which
is proportional to $\lambda^2/\chi$, is in fact inversely proportional
to the frequency. It is therefore the case that high-frequency waves
that have very short thermal relaxation times are isothermal, or
relaxed, and low-frequency waves are adiabatic, or unrelaxed. This
quite counter-intuitive situation is exactly the opposite of what one
has in the case of flexural vibrations of beams, and is probably the
best example of a system in which the relaxation time $\tau$
depends on the frequency $\omega$.

Nevertheless, as can be seen for example in the work of Chadwick and
Sneddon,\cite{chad} thermoelastic damping of longitudinal waves still
takes the form of a relaxation peak, with the isothermal and adiabatic
limits exchanged, and a with geometry independent characteristic
thermal relaxation time $\tau_l=\chi /c_l^2$ where $c_l$ is the speed,
or phase velocity, of longitudinal waves. Thus, the position of peak
thermoelastic damping for longitudinal waves is fixed and only depends
on the thermodynamic properties of the material as a function of
temperature. Some typical values for the thermal relaxation rates of
longitudinal waves in GaAs and Silicon are given in
Table~\ref{longTable}.

\section{Conclusions}
\label{conc}

We have established here that thermoelastic damping is a significant
source of dissipation down to the nanometer scale. We gave a simple
derivation of an exact expression \(qinv) for thermoelastic damping in
thin rectangular beams, compared this exact expression with Zener's
well known approximation \(zenerq), and examined the implication of
our result on micrometer- and nanometer-scale resonators. It is
interesting to note, as a consequence of our analysis [see
Figure~\ref{geometryplot} ($a$) and ($b$)], that for beams of constant
aspect ratio and constant temperature above a certain beam width the
quality factor increases {\it linearly\/} with the size of the
beam. This may provide a partial explanation for the linear increase
in dissipation as systems become smaller.

We have made a number of approximations and assumptions along the way
which we would like to summarize here:

1.~We have derived and solved the thermoelastic equations of a {\it
thin\/} beam undergoing {\it small\/} flexural vibrations. We should
not expect our result to strictly hold for beams with small aspect
ratios $L/b$. Nevertheless, we do expect to see the same kind of
behavior, showing a Debye-like dissipation peak, even at smaller aspect
ratios. We do not expect our result to hold for large amplitude
vibrations where the Euler-Bernoulli assumption is known to fail, and
where non-linear behavior begins to take over.

2.~In displaying the expected relaxation strengths
(Figure~\ref{strengthplot}) and damping curves
(Figure~\ref{geometryplot}) for GaAs and Silicon resonators with
various geometries we used experimentally-reported {\it bulk\/}
elastic and thermodynamic properties. According to recent moleculer
dynamics simulations of thin quartz beams by Broughton {\it et
al.}\cite{broughton} one is justified in using bulk properties down to
beam widths of about 10 micrometers. For smaller widths, quantities
such as Young's modulus change drastically. In any case, one should
consider the plots only as serving for illustrative purposes, in
particular because some of the thermodynamic properties, especially
below about 200$K$, vary from sample to sample.

3.~The theory of thermoelasticity is valid in a regime where thermal
phonons are diffusive and a temperature field can be defined
locally. One should expect to see deviations from this theory when the
phonon mean free path becomes comparable to the system size, or when
the relaxation rate of the phonons to their equilibrium Bose
distribution becomes comparable to the resonator's frequency. In the
first case, where the transport of thermal energy crosses over from
being diffusive to being ballistic one may expect to see thermal
relaxation times that are linear in the beam width (proportional to
$b/v$ where $v$ is the phonon velocity) instead of the diffusive
quadratic dependence $b^2/\chi$.  The second effect introduces an
additional dissipation mechanism which may be viewed as caused by the
``viscosity'' of the phonon gas as it relaxes to its equilibrium
state. This latter mechanism is called the ``Akhiezer
effect.''~\cite{akhi} We intend to explore these corrections to the
diffusive regime, as well as the fully ballistic phonon regime, in the
context of mesoscopic systems in a future publication.

\acknowledgments

We thank Michael Cross for many helpful conversations. We also thank
members of our group: Eyal Buks, Kamil Ekinci, Darrell Harrington, Raj
Mohanty, and Keith Schwab for ongoing discussion and
collaboration. We gratefully acknowledge support from DARPA MTO/MEMS
under grant DABT63-98-0012.


\appendix
\section{Where is the Lorentzian?}

Because Zener's approximation \(zenerq) using a single Lorentzian in
the variable $\omega\tau_{\rm z}$ is so good, it is illuminating to try to
identify this Lorentzian behavior in our own result \(qinv). To do so let
us expand both numerator and denominator of $Q^{-1}/\Delta_E$ in
powers of $\xi$,
\begin{eqnarray}\label{expansion}
{Q^{-1}\over\Delta_E} &=& 6 
{
{\frac{4}{5!}\xi^2 + \frac{8}{9!}\xi^6 +\ldots}
\over
{1+ \frac{1}{4!}\xi^4 + \frac{1}{8!}\xi^8 +\ldots}
} \nonumber \\[6pt]
&=& 6 {
{\frac{4}{5!}\xi^2 (1+ \frac{2\cdot 5!}{9!}\xi^4 
+ \frac{3\cdot 5!}{13!}\xi^8 +\ldots)}
\over
{1+ \frac{\xi^4}{4!} (1+ \frac{4!}{8!}\xi^4 
+\frac{4!}{12!}\xi^8 +\ldots)}
}\ .
\end{eqnarray}

One can clearly see that for small values of $\xi$, the two series in
parentheses each tend to 1, their ratio approaching 1 from above as
$\xi$ decreases. For large values of $\xi$, the leading 1 in the
denominator may be neglected, and one can show with a little bit of
effort that the ratio of the two series in parentheses tends to $5/4$
from below as $\xi$ increases. Defining ${\cal L}$ as the Lorentzian
\begin{equation}\label{lorentzian}
{\cal L}(\eta) = {\eta\over{1+\eta^2}}\ ,
\end{equation}
we see that for any value of $\xi$, $Q^{-1}/\Delta_E$ is bounded
between the two Lorentzians
\begin{equation}\label{bounds}
\frac{2\sqrt{6}}{5} {\cal L}\biggl({\xi^2\over\sqrt{24}}\biggr) \ \leq\ 
{Q^{-1}\over\Delta_E} \ \leq\ 
\frac{\sqrt{6}}{2} {\cal L}\biggl({\xi^2\over\sqrt{24}}\biggr) .
\end{equation}

Figure~\ref{universalplot} shows the universal thermoelastic damping
curve $Q^{-1}/\Delta_E$ in relation to these two bounding
Lorentzians. The inset shows the difference between Zener's
Lorentzian approximation, which in the above notation takes the form
\begin{equation}\label{ZenerL}
Q^{-1}_{\rm Z} = \Delta_E\ {\cal L}\biggl({\xi^2\over{\pi^2/2}}\biggr) ,
\end{equation}
and the exact result \(qinv). Note that $\pi^2/2\simeq4.935$ and
$\sqrt{24}\simeq4.899$ differ by less than $1\%$. It should be
emphasized that on the isothermal side of the peak (low frequencies)
the two expressions differ by less than $2\%$, which is the error
anticipated by Zener in keeping only the first term in his
expansion. On the adiabatic side of the peak (high frequencies) the
error increases in the first decade to as much as $15\%$, reaching
$20\%$ in the limit of infinite $\xi$.


\end{document}